\documentclass[epj]{svjour}
\usepackage{latexsym}
\usepackage[T1]{fontenc}
\usepackage[utf8]{inputenc}
\usepackage{amsmath}
\usepackage{amssymb}
\usepackage{graphicx}
\usepackage{grffile}
\usepackage{refstyle}
\usepackage[normalem]{ulem}
\usepackage{rotating}
\usepackage{xcolor}
\usepackage{multirow} 
\newcommand{\fss}{fastest switching stochastic on-off coupling }
\newcommand{\ag}{\textcolor{black}}
\begin{document}
\title{Comprehending deterministic and stochastic occasional uncoupling induced synchronizations through each other}
\subtitle{}
\author{Anupam Ghosh
\thanks{\emph{Present address:} Department of Aerospace Engineering, Indian Institute of Technology Madras, Chennai 600036, India.}%
\and Sagar Chakraborty}                     
\offprints{anupamghosh0019@gmail.com (A. G.)} 
%
\institute{Department of Physics,
	Indian Institute of Technology Kanpur,
	Uttar Pradesh 208016, India}

\date{}
%
\abstract{
In this paper, we \ag{numerically} study the stochastic and the deterministic occasional uncoupling methods of effecting identical synchronized states in low dimensional, dissipative, diffusively coupled, chaotic flows that are otherwise not synchronized when continuously coupled at the same coupling strength parameter. In the process of our attempt to understand the mechanisms behind the success of the occasional uncoupling schemes, we devise a hybrid between the transient uncoupling and the stochastic on-off coupling, and aptly name it the transient stochastic uncoupling---yet another stochastic occasional uncoupling method. Our subsequent investigation on the transient stochastic uncoupling allows us to surpass the effectiveness of the stochastic on-off coupling with very fast on-off switching rate. Additionally, through the transient stochastic uncoupling, we establish that the indicators quantifying the local contracting dynamics in the corresponding transverse manifold are generally not useful in finding the optimal coupling region of the phase space in the case of the deterministic transient uncoupling. In fact, we highlight that the autocorrelation function---a non-local indicator of the dynamics---of the corresponding response system's chaotic time-series dictates when the deterministic uncoupling could be successful. \ag{We illustrate all our heuristic results using a few well-known examples of diffusively coupled chaotic oscillators.}
} 

\authorrunning{Ghosh and Chakraborty}\titlerunning{Comprehending occasional uncoupling induced synchronization}
\maketitle
\section{Introduction} 

\ag{Synchronization---the coordinated motion of dynamical systems by dint of interactions among them---is espied in various real-life examples~\cite{strogatz03} like flock of birds, flashing of fireflies, insect swarms, school of fishes, etc. Among biological systems, synchronized circadian rhythms in daily life activity of animals and plants are among the first ones to be modelled using coupled oscillators~\cite{winfree67}. Furthermore, in other biology systems like chorusing of frogs~\cite{aihara08} and united firing of neurons~\cite{montbrio15}, synchronization is observed. Besides, technological applications in, say, an array of coupled lasers~\cite{winful90,roy94}, power grids' dynamics~\cite{motter13}, coupled Josephson junctions~\cite{wiesenfeld96}, and coupled thermoacoustic systems~\cite{pawar17}, we can detect the phenomenon of synchronization. Once the seemingly counterintuitive synchronization in interacting chaotic systems got established firmly~\cite{pc1990}, it has taken the centre stage in the research of synchronization phenomenon over the last three decades.} 
\ag{In coupled chaotic systems, different types of synchronization have been ascertained~\cite{pea1997,pikovsky01,bocc02,balanov08}: \emph{viz.}, phase synchronization, complete synchronization, lag synchronization, generalized synchronization, etc. Although, most of the literature on chaotic synchronization is about the coupled dissipative chaotic systems, the phenomenon has also been seen in time-delayed chaotic oscillators and chaotic coupled Hamiltonian systems. In Hamiltonian systems, the absence of any attractor---owing to the Liouville's theorem---leads to fundamentally different kind of synchronization called measure synchronization~\cite{hampton99}. The quantum counterpart of the measure synchronization~\cite{sur20} and extended measure synchronization in coupled quantum systems~\cite{qiu14} have also been reported. Furthermore, synchronization has been observed in laboratory experiments on coupled bosonic Josephson junctions~\cite{tian13m}, interacting ultra-cold atomic clouds~\cite{qiu15}, and interacting optomechanical systems~\cite{bemani17}.}

Although synchronization of chaotic dynamical systems is a well-documented and heavily investigated phenomenon in the field of nonlinear dynamics, it still has many surprising and ill-understood aspects---one of them being the synchronization brought about by the occasional uncoupling in diffusively coupled chaotic systems which fail to evolve in synchrony when continuously coupled~\cite{stojanovski96,zochowski00,cqh09,jeter15,sch15,li18}. An occasional uncoupling scheme of synchronization, by definition, means that the coupling between the chaotic oscillators---a drive and a driven, say---is repeatedly switched on and off  either stochastically, e.g., in the stochastic on-off coupling~\cite{jeter15} or deterministically, e.g., in the sporadic coupling~\cite{stojanovski96}, the intermittent coupling~\cite{zochowski00}, the on-off coupling~\cite{cqh09}, the transient uncoupling~\cite{sch15}, and the periodic coupling~\cite{li18}. Beside saving interaction costs owing to the comparatively reduced feedback between the participating chaotic oscillators, the occasional uncoupling has the advantage that it can impart synchronization without having to directly tune the coupling parameter. Of course, all such methods can be extended to the networks of the chaotic oscillators~\cite{hagberg08,chen10,kumar15,sch16,zhou16,jeter18,li18,chowdhury19}. \ag{It should also be pointed out that more detailed models of the diffusive coupling induced synchronization of nonlinear oscillators have been studied~\cite{marin07,bressloff17,gou17} where one explicitly uses partial differential equations to model diffusion. Interestingly, in such a system~\cite{bressloff17}, a variant of the stochastic on-off coupling has been implemented successfully.}

There is yet another similarity among most of the stochastic and the deterministic uncoupling induced synchronizations: there is no clear understanding of what the precise mechanism behind such a synchronization phenomenon is. As we discuss in this paper, neither the local analysis~\cite{tandon16,ghosh18} pin-pointing how the driven system's corresponding attractor's fraction that has locally contracting regions is affected by such uncoupling nor the idea of averaging~\cite{jeter15,ghosh18_2} that seeks to define a `renormalized' coupling parameter found after averaging the coupling strength over the on-off period of the coupling, completely explains the success or the failure of the occasional uncoupling across variety of chaotic systems. The concept of averaging especially fails---in both the relevant stochastic and deterministic schemes---when larger networks of oscillators are considered~\cite{jeter14,sch16} or when the coupling is switched on and off rather infrequently during the evolution of the systems. It is interesting to note that when the aforementioned averaged dynamics does not settle on to a synchronized state, there exist `windows of opportunity' when it is the infrequent or slow switching of coupling that can lead to synchronization. However, how to predict the existence of such windows in a diffusively coupled system is a very challenging ask---especially, for flows---although some convincing progress have been made for maps~\cite{golovneva17,porfiri17,jeter18}.
In view of the above, it is not unjustified that we believe that all hidden mechanisms behind all the uncoupling schemes must be heavily intertwined. In this paper, our goal is to shed more light on the phenomenon of the occasional uncoupling induced synchronization. Quite unexpectedly, we find that the success of stochastic uncoupling clears up why local analyses are not enough to understand deterministic schemes; additionally, the deterministic occasional uncoupling can be used to render the stochastic uncoupling even more effective. Such a symbiotic relationship between a deterministic phenomenon and its stochastic counterpart is, although not singular (e.g., the phenomenon of the vibrational resonance~\cite{landa_00}), quite rare. \ag{Although, in this paper, we do not make any mathematically rigorous statement regarding the working principles of the occasional uncoupling schemes, we do present heuristic insights following systematic case-wise studies of a few sets of coupled oscillators. In fact, the heuristic understanding helps us to propose an improved uncoupling scheme---transient stochastic scheme---discussed later in this paper.}
To this end, we discuss the \emph{fastest} stochastic uncoupling in Section~\ref{sec:ocs} and follow that up with a detailed Section~\ref{sec:revisit} scrutinizing the present state of understanding of the transient uncoupling induced synchronization. In Section~\ref{sec:tsu} we invent a new stochastic scheme, called \emph{transient stochastic uncoupling}, that is central to our idea of comprehending the deterministic and the stochastic occasional uncoupling induced synchronizations through each other. Section~\ref{sec:discussion} reiterates and discusses further the results of this paper. However, before anything else, we give an chronological eclectic review of the occasional uncoupling schemes in the immediately following section, Section~\ref{sec:ocs}.
\section{Occasional uncoupling schemes}
\label{sec:ocs}
Let us consider the case of two identical chaotic oscillators coupled diffusively and unidirectionally: 
\begin{subequations}
	\begin{eqnarray}
		\frac{d\mathbf{x}_1}{dt} &=& \mathbf{F(x}_1),\label{eq:1a}\\
		\frac{d\mathbf{x}_2}{dt}& =& \mathbf{F(x}_2) + \alpha  \chi (t, \mathbf{x}_1,\mathbf{x}_2) \sf{C}\cdot(\mathbf{x}_1 - \mathbf{x}_2).\label{eq:1b}
	\end{eqnarray}
	\label{eq:1}
\end{subequations}
Here $\mathbf{x}_1 (t)$ and $\mathbf{x}_2 (t)$ are respectively the states of the drive and the driven $d$-dimensional autonomous subsystems. The matrix $\sf{C}$ is the $d \times d$ coupling matrix, and $\alpha$ is a scalar that measures the coupling strength.  $\chi(t, \mathbf{x}_1,\mathbf{x}_2)$ is a scalar function that can take only two discrete values---$0$ and $1$ . Thus, one may note that technically it is the $\chi$ which makes the diffusive coupling \emph{occasional}. Depending on what form of $\chi$ is chosen, one can have a particular form of occasional uncoupling scheme. These schemes of synchronization can be broadly classified into deterministic and stochastic schemes. In what immediately follows, we give an eclectic and chronological review of such schemes having a predefined form of $\chi$. Subsequently, we introduce the \fss that is crucial for the results we arrive at in this paper.

The sporadic coupling~\cite{stojanovski96,stojanovski97} is a deterministic occasional coupling scheme where the drive sends an instantaneous signal after every constant time interval  ($\Delta t$) to the response subsystem. Mathematically speaking, 
\begin{equation}
	\chi(t, \mathbf{x}_1,\mathbf{x}_2)=\chi_{\Delta t}(t) := \sum_{n}\delta(t- n\Delta t),\quad n\in\mathbb{N};
\end{equation}
where $\delta(t)$ is the Dirac delta function normalized to unity. 

Similar to the sporadic coupling, in the intermittent coupling~\cite{zochowski00}, the drive sends an instantaneous signal to the response subsystem whenever the drive's state is on the $(d-1)$-dimensional Poincar\'e section $\mathbb{P}_I (\mathbf{x}_1)$, i.e.,
\begin{equation}
	\hspace*{-0.3 cm}
	\quad\chi(t, \mathbf{x}_1,\mathbf{x}_2)=\chi_{\mathbb{P}_I} (t) := \begin{cases}
		1,  \, \text{for}\, t = T_0, \mathbf{x}_1(T_0) \in \mathbb{P}_I (\mathbf{x}_1);\\
		0,  \, \text{otherwise}.
	\end{cases}
\end{equation}

The synchronization can be brought about even with stochastic uncoupling: in the stochastic on-off coupling~\cite{belykh04}, after each $t = n\tau$ a random number is called from a uniform distribution of real numbers in the interval $[0,1]$, and the drive oscillator couples to the driven with probability $p$. More compactly, we can say that in this scheme: 
\begin{equation}
	\hspace*{-0.3 cm}
	\chi(t, \mathbf{x}_1,\mathbf{x}_2)=\chi_{(\tau, p)} (t) := \begin{cases}
		1 \, \text{with probability $p$ for}\\
		\phantom{1,} \, n\tau < t \leq (n+1)\tau;\\
		0 \, \text{with probability $1-p$ for}\\
		\phantom{1,} \, n\tau < t \leq (n+1)\tau.
	\end{cases}
	\label{eq:sof}
\end{equation}
If $\tau$ is small compared to the system timescale ($T_s$), the coupling is called the fast switching stochastic on-off coupling. However, for larger values of $\tau$---i.e., when $\tau$ and the system timescale have same order of magnitude---there exists the so-called `windows of opportunity'~\cite{jeter15} that help to pick a combination of $\tau$ and $p$ leading to synchronization. This stochastic scheme is called the slow switching stochastic scheme.

Somewhat along the similar line, if one defines
\begin{equation}
	\hspace*{-0.3 cm}
	\chi(t, \mathbf{x}_1,\mathbf{x}_2)= \chi_{(T, \theta)} (t) := \begin{cases}
		1,  \, nT < t \leq (n+\theta)T;\\
		0,  \, (n + \theta)T < t \leq (n+1)T,
	\end{cases}
	\label{eq:dof}
\end{equation}
where $T$ and $\theta$ are fixed, the resulting occasional uncoupling scheme would be deterministic. This is known as the on-off coupling~\cite{cqh09}. It is interesting to note that the plot of $\chi_{(T, \theta)} (t) $~vs~$t$ can be seen as a particular realization of $\chi_{(\tau, p)} (t)  $~vs~$t$ plot whenever $\theta=m_1/(m_1+m_2)$ ($m_1,m_2\in\mathbb{N}$).

While the aforementioned on-off coupling is temporal in nature, i.e., $\chi$ depends only on $t$ explicitly, the transient uncoupling~\cite{sch15,sch16,tandon16} is spatial in nature such that $\chi$ depends only on $\mathbf{x}_2$ and not $t$ explicitly. In other words, only when the trajectory of the driven oscillator is within a particular (predefined) region of phase space,  both the oscillators are coupled. Therefore,
\begin{equation}
	\chi(t, \mathbf{x}_1,\mathbf{x}_2)=\chi_{_{\mathbb{A}}} (\mathbf{x}_2) := \begin{cases}
		1,  \,  \mathbf{x}_2 \in \mathbb{A};\\
		0,  \,  \mathbf{x}_2 \notin \mathbb{A},
	\end{cases}
	\label{eq:chitus}
\end{equation}
where $\mathbb{A} \subseteq \mathbb{R}^d$ is the predefined region in the phase space of the response oscillator.

Last but not the least, in the periodic coupling~\cite{li18}, the term $\chi(t, \mathbf{x_1},\mathbf{x}_2)$ in Eq.~\ref{eq:1}b is a function of $t$ only and is defined as:
\begin{equation}
	\chi(t, \mathbf{x}_1,\mathbf{x}_2)= \chi_\omega (t) :=  \sin\omega t+1 .
\end{equation}
Here the function $\chi_\omega (t)$ does not take binary values but changes sinusoidally with a frequency $\omega$. Also, in contrast to the sporadic coupling and the intermittent coupling, in this scheme, the coupling term is mostly active as the system evolves; the uncoupling happens rather sporadically/intermittently whenever {$t=(2n-0.5)\pi/\omega$}.

Finally, we introduce a particular form of fast switching stochastic on-off coupling, viz., the fastest stochastic on-off coupling scheme. Let us consider that the state of the coupled system is discretely updated---as is expected under any numerical algorithm---after every time interval $h$ ($\ll T_s$; $T_s$ being the system's smallest timescale) in accordance with the equations of motion (Eq.~\ref{eq:1}) governing it. The fastest possible fast switching stochastic on-off coupling (Eq.~\ref{eq:sof}) would be in action when $\tau=h$. We term such coupling, fastest switching stochastic on-off coupling, and the corresponding $\chi$ can be conveniently written as
\begin{equation}
	\label{eq:new_condi}
	\chi(t, \mathbf{x}_1,\mathbf{x}_2)= \chi_{(\xi, q)}(t) := \begin{cases}
		1  \text{ for }\xi(t) \geq q,\\
		0  \text{ for }\xi(t) < q;
	\end{cases}
\end{equation}
where $q\in(0,1)$ is some fixed threshold number and $\xi(t)\in[0,1]$ is a uniform random variable. Numerical studies (see~\ref{sec:appendix}) show that the \fss is better compared to other mentioned occasional uncoupling schemes as far as synchronization at higher coupling strength is concerned. This effectiveness of the \fss may be traced to the idea of an effective coupling strength~\cite{jeter15}, explained in~\ref{sec:appendix}. 
\ag{It is motivating to note that the occasional uncoupling is not merely a theoretical exercise in chaotic synchronization; various experimental systems~\cite{bocc00,bocc02} are amenable to it. For example, consider thermoacoustic instability---detected in various thermoacoustic systems~\cite{juniper18}---that is are harmful to various real-life combustors used in industrial furnaces, ramjets, and rockets~\cite{fisher09}. The horizontal Rijke tube~\cite{rijke59} is one such example of thermoacoustic systems. The Rijke tube is a cylindrical glass tube with a wire mesh within it, and both the ends of the tube are open~\cite{rijke59}. In certain experiments, two such tubes are coupled through another tube that is called coupling tube; and as a result, phase and generalized synchronizations are observed between heat release rate and acoustic pressure~\cite{thomas18}. It is straightforward to employ the on-off coupling scheme in this set-up by adding a switch in the coupling tube such that only when the switch is on, there is coupling between the tubes. Thus, one may be able to control the aforesaid instability through the occasional uncoupling leading to synchrony when the continuous coupling fails. As another example, consider a driven diode resonator which generally consists of a p-n junction, an inductor, and a sinusoidal current source, that are connected in series~\cite{hunt91,bocc00}. The p-n junction gets feedback from the current source. The initial chaotic dynamics of the diode resonator switches to periodic orbits because of the usually employed periodic occasional feedback. One may arrange for the feedback current to be sent only when its amplitude is within a predefined window. Similarly, there exist many other examples where one may implement the occasional uncoupling in, say, control the stability of laser~\cite{roy94}, chemical reactions~\cite{patrov92}, biological systems~\cite{garfinkel92}, etc.}
\section{Revisiting Transient Uncoupling}
\label{sec:revisit}
\begin{figure*}[t]
	\hspace*{-0.5cm}
	\includegraphics[width=19cm,height=7.4cm, keepaspectratio]{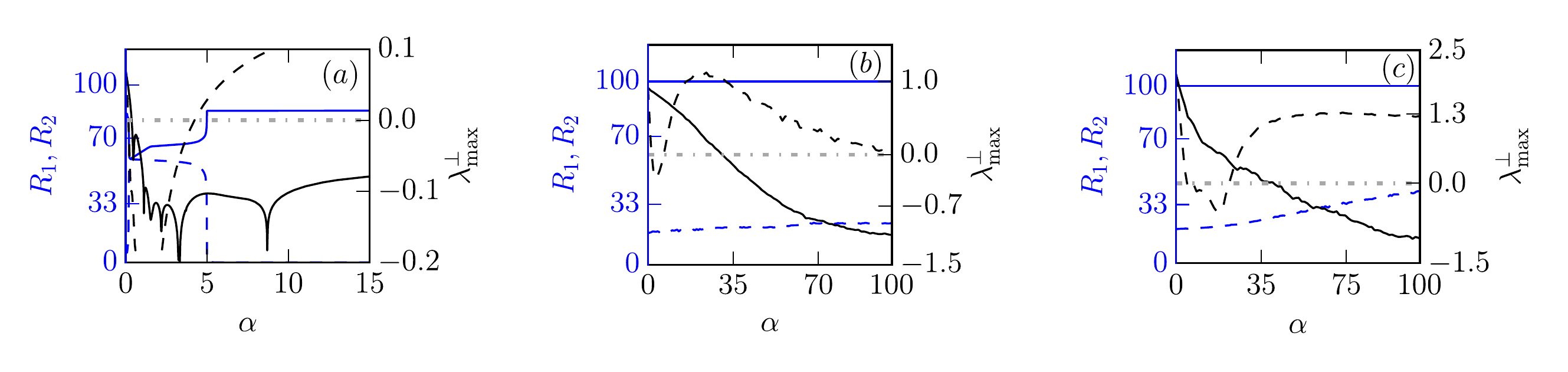}
	\caption{\textit{(Color online)} \textbf{Measures of locally contracting orbits may not indicate optimal coupling region:} $R_1$ (blue dashed curve) and $R_2$ (blue solid curve) along with the maximum conditional Lyapunov exponent, $\lambda_{\rm max}^\perp$, corresponding to the continuous coupling (black dashed curve) and the transient uncoupling (black solid curve) are plotted with coupling strength $\alpha$ for the coupled R\"ossler oscillators (subplot a), the coupled Lorenz oscillators (subplot b), and the coupled Chen oscillators (subplot c). $R_1$ and $R_2$ together are unambiguously able to detect the observed synchronized states at higher $\alpha$ only for the coupled R\"ossler oscillators. Note the sharp transitions in subplot (a). Synchronized states, i.e., $\lambda_{\rm max}^\bot<0$, is observed for $\alpha \gtrsim 30.5$ and $\alpha \gtrsim 41$ for the coupled Lorenz and the coupled Chen oscillators respectively. In both the cases, $R_2$ does not change with $\alpha$, thereby becoming a non-participant in detecting the optimal coupling region. The other parameter, $R_1$, has a small variation with $\alpha$ in (c), but---because of having no sharp transition---is unable to mark the transition to the synchronized state at $\alpha \approx 41$. In (b), $R_1$, similar to $R_2$, is almost constant as $\alpha$ varies, and hence is yet again unable to tell whether synchronization could be induced. The grey dot-dashed horizontal line, $\lambda_{\rm max}^\bot=0$, is merely an aid to the eyes.}
	\label{fig:r1_r2}
\end{figure*}

\ag{Having seen the different types of the occasional uncoupling schemes, we now turn our attention towards systematic numerical experiments. To this end, throughout this paper, we work exclusively with coupled three dimensional autonomous flows, i.e., with Eq.~\ref{eq:1} where $d =3$.} The transient uncoupling scheme is more enigmatic to understand. Of particular interest is the question what predefined region, $\mathbb{A}$ (Eq.~\ref{eq:chitus}), to optimally choose such that the transient uncoupling induces synchronization. It seems tempting and intuitive that, since the eigenvalues of the Jacobian of the transverse dynamics characterize whether an orbit at a \ag{point (henceforth, called phase point) in phase space of the error dynamics} is locally diverging away from another nearby orbit, an optimal coupling region, $\mathbb{A}$, should ideally be chosen in such a manner that at most of its phase points either the corresponding maximum of the real parts of the eigenvalues is negative or the magnitude of the maximum positive real part is lesser than that of the minimum negative real part. 

Technically, the aforementioned idea is as follows~\cite{ghosh18}: Let $\Lambda_{\rm max} \geq \Lambda_{\rm mid} \geq \Lambda_{\rm min}$ be the real parts of the eigenvalues of the local Jacobian of the response oscillator. We define two parameters:
\begin{eqnarray}
	R_1 &:=& 100 \times \frac{N_{-}}{{N_{-}+N_{+}}},\\
	R_2 &:=& 100\times\frac{\textrm{Fraction of $N_+$ with }|\Lambda_{\rm max}| < | \Lambda_{\rm min}| }{N_+},\qquad
\end{eqnarray}
where $N_-$ and $N_+$ are the total numbers of phase points with negative and positive $\Lambda_{\rm max}$ respectively. It is obvious that larger values of $R_1$ {or} $R_2$ should indicate the synchronized state. 

We see below that although, this idea meets with success in the case of the R\"ossler oscillator, unfortunately, it may not be able to explain the transient uncoupling induced synchronization in other diffusively coupled chaotic oscillators. For simplicity we work with three dimensional systems---the R\"ossler oscillator, the Lorenz oscillator, and the Chen oscillator. In all the three cases we chose the form of $\mathbb{A}$ as follows:
\begin{equation}
	\label{eq:tus_condi}
	\chi_{\mathbb{A}}(\mathbf{x}_2) = \begin{cases}
		1  \text{ if } |(\mathbf{x}_2)_i - x_{i0}| \leq \Delta,\\
		0  \text{ otherwise}.
	\end{cases}
\end{equation}
Here $x_{i0}\in\mathbb{R}$, $\Delta\in\mathbb{R}^+$, and $i \in \{1,2,3\}$ are chosen suitably case by case: $(i,x_{i0},\Delta)$ for the  $x$-coupled R\"ossler oscillators (Fig.~\ref{fig:r1_r2}a), the $z$-coupled Lorenz oscillators (Fig.~\ref{fig:r1_r2}b), and the $z$-coupled Chen oscillators (Fig.~\ref{fig:r1_r2}c) is respectively $(1,1.20, 4.16),\,(3,25, 1),$ and $(3,26.5, 5)$. Please refer to Table~\ref{table:2} for the explicit mathematical equations for the oscillators.

A critical and close inspection of the numerical results illustrated in Fig.~\ref{fig:r1_r2} unequivocally speaks volumes for the fact that $R_1$ and $R_2$ are not suitable indicators for choosing optimal coupling region across dynamical systems. It is clear~(Fig.~\ref{fig:r1_r2}) from the plots of $\lambda_{\rm max}^\bot$~vs.~$\alpha$ for all the three systems, the transient uncoupling presents a major improvement over the continuous coupling because in the former, the systems do get synchronized at much higher values of the coupling parameters. In the coupled R\"ossler systems, the value of $R_1$ is quite high (approximately $60$) and as soon as $R_1\rightarrow0$ near the upper threshold of synchronization, $R_2$ jumps to a high value (approximately $80$) that is constant with increasing $\alpha$~(Fig.~\ref{fig:r1_r2}a) . Hence, this is consistent with the success of the transient coupling employed on the system. Fig.~\ref{fig:r1_r2}b, however, renders $R_1$ and $R_2$ hopeless as far as any prediction about the success of the transient uncoupling in the case of the coupled Lorenz systems is concerned; $R_1$~vs.~$\alpha$ remains almost constant and non-zero (approximately $25$). In this context, we also note that $R_2$ is almost at its maximum possible value ($100$) and constant---a feature also observed in the coupled Chen systems~Fig.~\ref{fig:r1_r2}c. In the coupled Chen systems, although $R_1$ in monotonically increasing, there is no sharp transition marking the upper threshold for the coupling parameter resulting in marginally synchronization state (i.e., $\lambda_{\rm max}^\bot=0$). {Therefore, no sharp variation is observed either in $R_1$ or in $R_2$ when there is transition from the desynchronized state to the synchronized state for coupled Lorenz and coupled Chen oscillators; which is observed for coupled R\"ossler oscillators.} This makes predicting whether the uncoupling region under test is optimal or not. 

In view of the above we have no other choice but to conclude that local analysis~\cite{ghosh18} proposed to identify optimal coupling region definitely is not the complete story behind why the transient uncoupling works the way it does in any arbitrary system. Thus, we again come back to the central question of the paper why and when an occasional uncoupling succeeds. 
\section{The transient stochastic uncoupling}
\label{sec:tsu}
It is intriguing to note that both random and deterministic occasional uncoupling schemes lead to synchronization when continuous coupling fails to do so. While one would have intuitively thought that understanding the mechanism behind the deterministic schemes is easier, it seems that the \fss is understood much more straightforwardly than the transient uncoupling. However, one should not be misled to think that the stochastic on-off coupling is fully understood; if the switch is not fast enough, there is no known general mechanism behind why the coupling scheme be successful. In summary, we emphasize that the general mechanism behind the effectiveness of almost every occasional uncoupling scheme in bringing about synchronization is far from being understood. This problem is, of course, very difficult---and in fact, may not have a unified solution---because different chaotic systems are bound to have different solutions; and to the best of our knowledge, we do not know if there exists any rigorous universality class of chaotic flows that are known to follow same mechanism of synchronization. Thus, given the enormity of the problem, we ask a few pertinent questions that may help us to understand better why occasional uncoupling---whether random or deterministic---works so effectively.
\subsection{The Questions}
\label{sec:question}
From the discussion in the immediate section, it is crystal clear that the working principle of the transient uncoupling  is not always understood through the information of local stability of the phase points of the system's phase trajectory. To be specific, while the mechanism based on this information explains the success of the uncoupling for the R\"ossler system, it fails for other systems. Consequently, a question arises: \emph{What properties of a system must be known to be able to prescribe when uncoupling may induce synchronization?}

Again, we recall that the \fss is an extremely effective uncoupling scheme applicable across various chaotic systems. It results in the synchronization even at those higher values of the coupling strength, $\alpha$, at which the continuous coupling fails. Since, by construction, the stochastic on-off coupling can not be switched faster than the fastest switching stochastic on-off coupling, we find the following question thought-provoking: \emph{ How can, if at all, the \fss be bettered such that the resulting stochastic on-off coupling scheme induces synchronization at those large coupling strengths at which even the \fss fails?}

In the rest of this section, we present a novel idea to answer these questions.
\begin{figure}[htpb!]
	\hspace*{1.2 cm}
	\includegraphics[width= 40 cm,height=12.0cm, keepaspectratio]{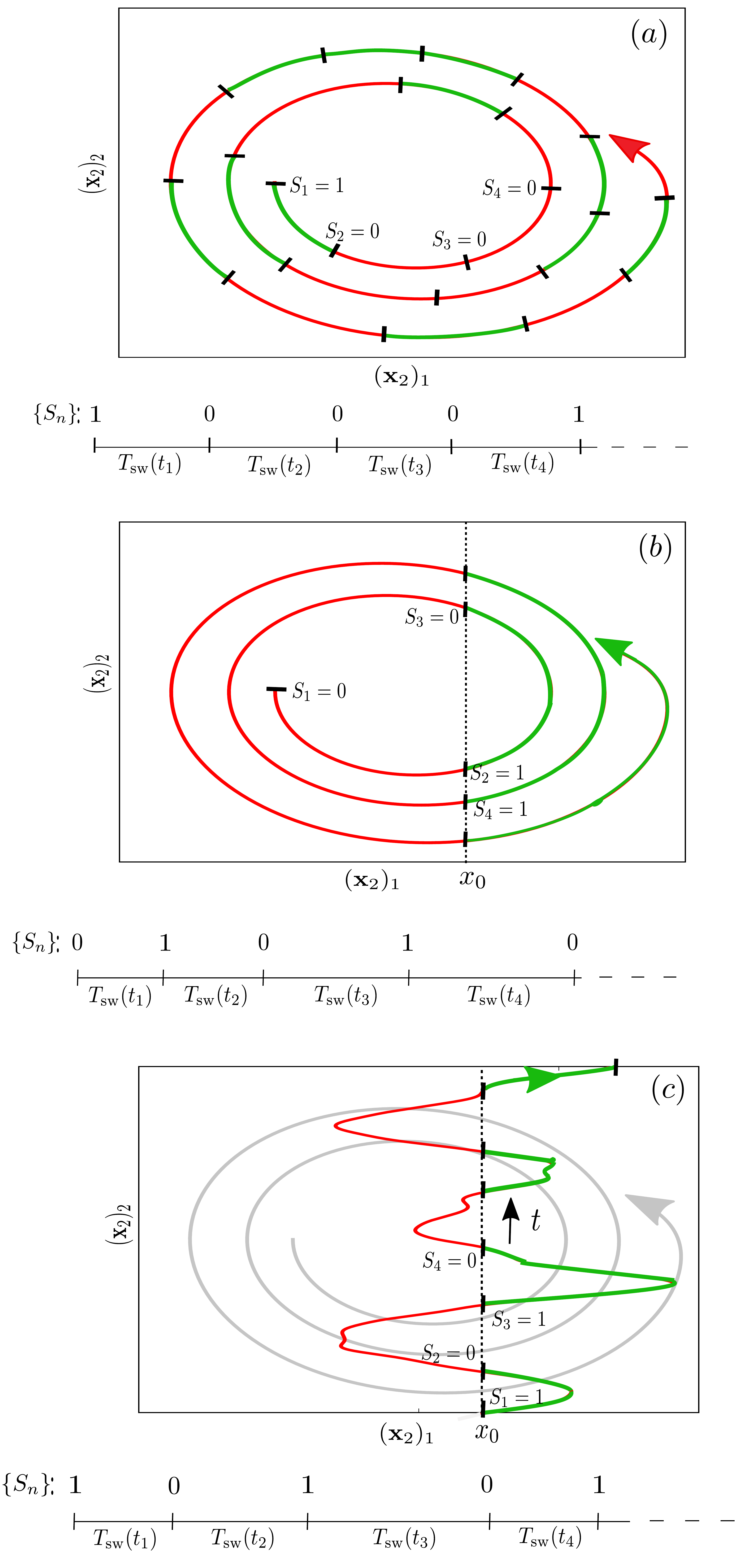}
	\caption{\textit{(Color online)} \textbf{Transient stochastic uncoupling is a hybrid between transient uncoupling and stochastic on-off coupling:} The schematic diagrams above illustrate the definitions of (a) stochastic on-off coupling, (b) transient uncoupling, and (c) transient stochastic uncoupling. In subplot (a), $T_{\rm sw}$, the time-interval after which coupling is turned on ($S_n=1$) or off ($S_n=0$), is constant but the sequence $\{{S}_n\}_{n\in\mathbb{N}}$ is collection of randomly chosen $0$'s and $1$'s. The red and the green regions of the representative attractor respectively indicate over which fractions of the attractor the coupling is inactive or active. $T_{\rm sw}(t_k):=t_{k+1}-t_k$ is the time elapsed between the events $S_k=0\,{\rm or }\,1$ and $S_{k+1}=0\,{\rm or }\,1$. In subplot (b), $T_{\rm sw}$ is time dependent but $\{{S}_n\}_{n\in\mathbb{N}}$ is a deterministic sequence; however, $T_{\rm sw}$ depends on time implicitly through the chaotic time-series, say $(\textbf{x}_2)_1$, corresponding to the attractor. In subplot (c), $T_{\rm sw}$ is time dependent but $\{{S}_n\}_{n\in\mathbb{N}}$ is a deterministic sequence; but, $T_{\rm sw}$ depends on time implicitly through a stochastic time-series which has same autocorrelation as $(\textbf{x}_2)_1$. This time series, plotted with time t, is shown as the red-green curve against the gray silhouette of the attractor.}
	\label{fig:sub}
\end{figure}	
\subsection{The Idea}
\label{sec:idea}
We start with promising speculation that since the transient uncoupling effects synchronization, one can probably include its salient features into the stochastic on-off coupling to improve the latter. To make this idea more concrete, we note two important points: firstly, the two schemes of the occasionally uncoupling under consideration are inherently different in nature as one is stochastic (random) and the other is deterministic. 

Nevertheless, the schemes are somewhat similar in the sense that in both the cases, the coupling is turned on or off after some time interval ($T_{\rm sw}$, say)---either fixed ($\tau$ in the stochastic on-off; see Eq.~\ref{eq:sof}) or time-dependent (in the transient uncoupling)---repeatedly as the system evolves. In other words, for each of the schemes there is a particular time-series, a sequence, $\{{S}_n\}_{n\in\mathbb{N}}$, of binary values (say, $0$ indicating uncoupling and $1$ indicating coupling) specified at $t=0$ and then after every subsequent elapsed time interval, $T_{\rm sw}$. For the stochastic on-off, $T_{\rm sw}$ is time independent but $\{{S}_n\}_{n\in\mathbb{N}}$ is a random sequence,  whereas for the transient uncoupling scheme, $T_{\rm sw}$ is time dependent but $\{{S}_n\}_{n\in\mathbb{N}}$ is a deterministic sequence; however, the time dependence of $T_{\rm sw}$ is not simple as it depends on time implicitly through the chaotic time-series of one of the variables of the driven subsystem. 

With this in mind, we back ourselves to the conjecture that if we invent a stochastic on-off scheme such that $T_{\rm sw}$ depends on time implicitly through the stochastic time-series that, up to some extent, is statistically identical to the aforementioned chaotic time-series. The new type of stochastic on-off scheme would inherit the pros of the transient coupling and may turn out to be even better than the fastest switching stochastic on-off coupling. To keep things simpler, here we may keep $\{{S}_n\}_{n\in\mathbb{N}}$ deterministic. Now we implement this idea.
\subsection {The Method: Phase Randomized Time-Series}
\label{sec:method}
Let us say that we have two diffusively coupled identical subsystems (Eq.~\ref{eq:1}) on which the transient uncoupling has been employed. For the sake of convenience and without any loss of generality, we assume that $\mathbb{A}$ in Eq.~\ref{eq:chitus} is such that
\begin{equation}
	\label{eq:tus_good}
	\chi(\mathbf{x}_2) = \begin{cases}
		1  \text{ for } (\mathbf{x}_2)_i \geq x_0,\\
		0  \text{ for } (\mathbf{x}_2)_i < x_0,
	\end{cases}
\end{equation}
for an $i\in\{1,2,3\}$ chosen conveniently and $x_0$ is an appropriately chosen real number.

For $(\mathbf{x}_2)_i$ under consideration, as is indispensable for any numerical algorithm, the evolution of the system yields a sequence $\{{s}_n\}_{n=0}^{N-1}$ of points sampled uniformly.
Let $\tilde{s}_k$ be the discrete Fourier transformation of $\{{s}_n\}_{n=0}^{N-1}$:
\begin{equation}
	\tilde{s}_k = \frac{1}{\sqrt{N}} \sum_{n=0}^{N-1} s_n \exp\left(-\sqrt{-1}\frac{2\pi n k}{N}\right),
\end{equation}
where $N$ is the total number of elements of $s_n$. Since $\{{s}_n\}_{n=0}^{N-1}$ is real sequence, $\{{s}_k\}_{k=0}^{N-1}$ is a complex sequence in general. We pick random phases $\phi_k$ from a uniform distribution whose range is $[0, 2\pi]$ while following the constraint, $\phi_{N-k} = -\phi_k$. Subsequently, we define
\begin{equation}
	\tilde{s}^\prime_k :=\tilde{s}_k \exp(-\sqrt{-1}\phi_k),
\end{equation}
for all $k$ except for $k=0$ in case of odd $N$ and except for $k=0\,{\rm and}\, N/2$ in case of even $N$.
The inverse Fourier transformation yields:
\begin{equation}
	{s}^\prime_n = \frac{1}{\sqrt{N}} \sum_{k=0}^{N-1} \tilde{s}^\prime_k \exp\left(\sqrt{-1}\frac{2\pi n k}{N}\right).
	\label{eq:surro}
\end{equation}
By construction, $\{{s}^\prime_n\}_{n=0}^{N-1}$ is a real but random sequence; and has the power spectrum and the autocorrelation function~\cite{maiwald08,lancaster18} identical to that of $\{{s}_n\}_{n=0}^{N-1}$.  Technically speaking, we have thus created a surrogate data set, using the phase randomization technique~\cite{kantz04}.

We are now fully equipped to propose a variant of occasional uncoupling such that $\chi$ in Eq.~\ref{eq:1}b is defined as
\begin{equation}
	\label{eq:tus_good1}
	\chi(t, \mathbf{x}_1,\mathbf{x}_2)=\chi(s'_n) := \begin{cases}
		1  \text{ for } s'_n \geq x_0,\\
		0  \text{ for } s'_n < x_0.
	\end{cases}
\end{equation}
We aptly call the resulting occasional uncoupling scheme: the \emph{transient stochastic uncoupling}. We observe that it is a stochastic on-off scheme such that $T_{\rm sw}$ depends on time implicitly through the stochastic time-series, $\{{s}^\prime_n\}_{n=0}^{N-1}$, that is statistically identical to the chaotic time-series, $\{{s}_n\}_{n=0}^{N-1}$, as far as the autocorrelation and the power spectrum are concerned. Fig.~\ref{fig:sub} schematically illustrates the definition of the transient stochastic uncoupling.

\subsection{The Numerical Results}
\label{sec:result}
\begin{figure*}[htpb!]
	\hspace*{0.7 cm}
	\includegraphics[width=50 cm,height=12cm, keepaspectratio]{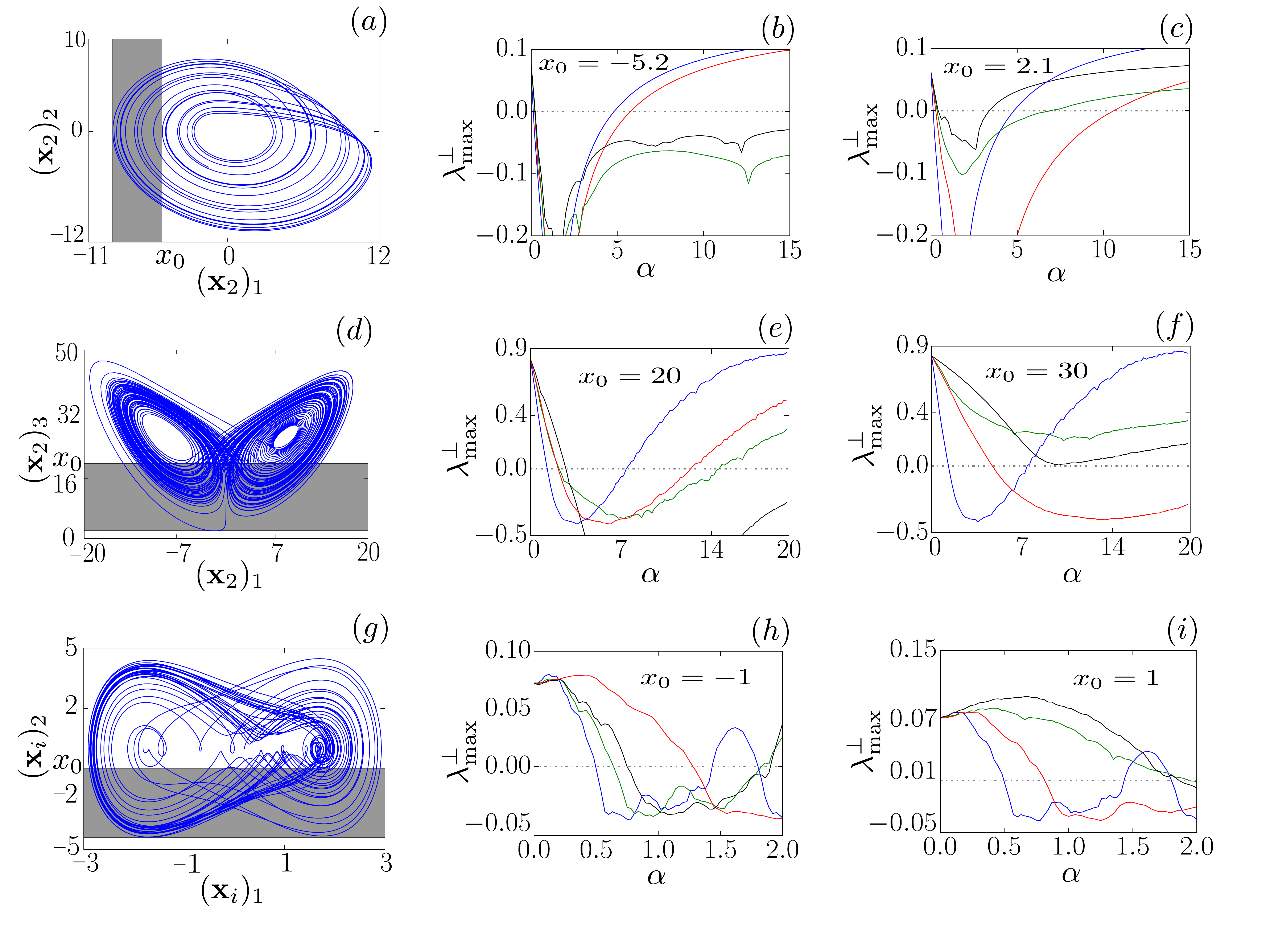}
	\caption{\textit{(Color online)} \textbf{Transient stochastic uncoupling can outperform \fss and the importance of autocorrelation:} Continuous coupling and three different occasional uncoupling methods are applied to the coupled R\"ossler oscillators (top row), the coupled Lorenz oscillators (middle row), and the coupled Duffing oscillators (bottom row). In each row, the grey-shaded regions of the projected phase portraits of the respective coupled systems depict the region where the coupling is inactive (subplot~a, d, and g). In subplots (b), (e), and (h), we observe that the transient uncoupling (black curve) and the transient stochastic uncoupling (green curve) lead to the synchronized states for higher $\alpha$ values compared to what the \fss (red curve) can achieve. Other subplots (c), (f), and (i) show that the \fss is better than the transient uncoupling, and hence than the transient stochastic uncoupling (sharing the same autocorrelation function with the transient uncoupling case), for some different choices of the coupling regions. In all six $\alpha$ vs $\lambda_{\rm max}^\perp$ subplots, the blue curves are for the continuous coupling.}
	\label{fig:tus}
\end{figure*}
Lets employ the transient stochastic uncoupling---along with the continuous coupling, the transient uncoupling, and the fast switching stochastic on-off---on the three coupled oscillators: the R\"ossler oscillator, the Lorenz oscillator, and the forced Duffing oscillator. It is worth commenting that most of the occasional uncoupling schemes are reported~\cite{zochowski00,belykh04,cqh09,chen10,sch15,li18} with the R\"ossler oscillator as the example and interestingly, as one tries to apply them to other systems, they more often than not fail (see Table~\ref{table:2}). This fact has motivated us to work with the aforementioned there different chaotic oscillators, including a non-autonomous one.
Before we proceed further, we remark that we are going to compare the four different schemes which include both deterministic and stochastic schemes, we put them on similar footing by picking the random numbers for the \fss from a uniform distribution within range: $[\min\left( \{{s}_n\}_{n=0}^{N-1} \right), \max\left( \{{s}_n\}_{n=0}^{N-1} \right)]$, which is size of the corresponding attractor along the coordinate that is supposed to generate $\{{s}_n\}_{n=0}^{N-1}$. {Therefore, the corresponding form of $\chi(t, \mathbf{x}_1,\mathbf{x}_2)$ (cf. Eq.~\ref{eq:new_condi}) may be written as: 
\begin{equation}
\label{eq:new_condi1}
\chi(t, \mathbf{x}_1,\mathbf{x}_2)= \chi_{(\xi, x_0)}(t) := \begin{cases}
	1  \text{ for }\xi(t) \geq x_0,\\
	0  \text{ for }\xi(t) < x_0;
\end{cases}
\end{equation}
the random numbers, $\xi(t)$, are picked from the aforementioned uniform distribution.}

First, we take unidirectionally $x$-coupled R\"ossler oscillators and investigate synchronization for the coupling strengths $\alpha \in [0,15]$. Here, the sequence $\{{s}_n\}_{n=0}^{N-1}$ is sampled from the $(\mathbf{x}_2)_1$-coordinate. We observe as depicted in Fig.~\ref{fig:tus}b, on using $x_0=-5.2$, i.e., $\mathbb{A}=\{\mathbf{x}_2\in\mathbb{R}^3|(\mathbf{x}_2)_1 \geq x_0\}$, the \fss leads to the synchronization (negative maximum conditional Lyapunov exponent) for a maximum value of $\alpha \approx 5.8$. In contrast, the transient uncoupling and the transient stochastic uncoupling results in the synchronized states even for $\alpha > 5.8$.  However, in the case of $x_0 = 2.1$ (Fig.~\ref{fig:tus}c), the \fss is better than the transient uncoupling and the transient stochastic uncoupling in terms of synchronizing at larger $\alpha$. 

Secondly, we consider unidirectionally $z$-coupled Lorenz oscillators~\cite{lor1963} with parameters $\sigma =10, r = 28,$ and $b = 8/3$ (see Table~\ref{table:2}). While employing the transient uncoupling and transient stochastic uncoupling, the coupling is active in the set  $\mathbb{A}=\{\mathbf{x}_2\in\mathbb{R}^3|(\mathbf{x}_2)_3 \geq x_0\}$. Similar to the case of the coupled R\"ossler oscillators, we observe that for some values of $x_0$ (e.g., $x_0 = 20$), the transient uncoupling and the transient stochastic uncoupling return synchronized state at larger $\alpha$ (Fig.~\ref{fig:tus}e) compared to the fast switching stochastic on-off coupling; whereas the converse is true for some other $x_0$ (e.g., $x_0 = 30$) as seen in Fig.~\ref{fig:tus}f. 

Finally, let us focus on bidirectionally coupled Duffing oscillators~\cite{stefanski07,jeter15}:
\begin{subequations}
	\begin{eqnarray}
		\dot{x}_i &=&  y_i, \\
		\dot{y}_i &=&x_i^3 -h y_i + q \sin(bt) + \alpha \chi(y_i) \cdot (x_j- x_i),	
	\end{eqnarray}\label{eq:duffing}
\end{subequations}
where $i,j = 1,2$ with $i \neq j$; and the parameters $b$, $h$, and $q$ are $1.0$, $0.01$, and $5.6$ respectively. Here, the sequence  $\{{s}_n\}_{n=0}^{N-1}$ is generated by either $y_1$ or $y_2$. Further, for the stochastic schemes---the \fss and the transient stochastic uncoupling---we picked the random numbers from a uniform distribution with boundaries $[\min\left( \{{s}_n\}_{n=0}^{N-1} \right), \max\left( \{{s}_n\}_{n=0}^{N-1} \right)] = [-4.54, 4.51]$. Here, $\mathbb{A}=\{\mathbf{x}_2\in\mathbb{R}^2|(\mathbf{x}_2)_2 \geq x_0\}$ with $x_0 = -1.0$ and $1.0$. In this system, let us concentrate on the lower threshold of synchronization, i.e., the value of the coupling parameter below which desynchronization happens. It is satisfying to note that the conclusions drawn for the R\"ossler and the Lorenz attractors applicable to this case also: both the transient stochastic uncoupling and the transient uncoupling  are either better ($x_0=-1$; Fig.~\ref{fig:tus}h) or worse ($x_0=1$; Fig.~\ref{fig:tus}i) than the fastest switching stochastic on-off coupling. The word `better' means that the corresponding occasional uncoupling starts imparting synchronization from comparatively lower value of $\alpha$

\subsection{The Conclusions}
\label{sec:conclusion}
\begin{figure*}[t]
	{
		\includegraphics[width=18 cm,height=18 cm, keepaspectratio]{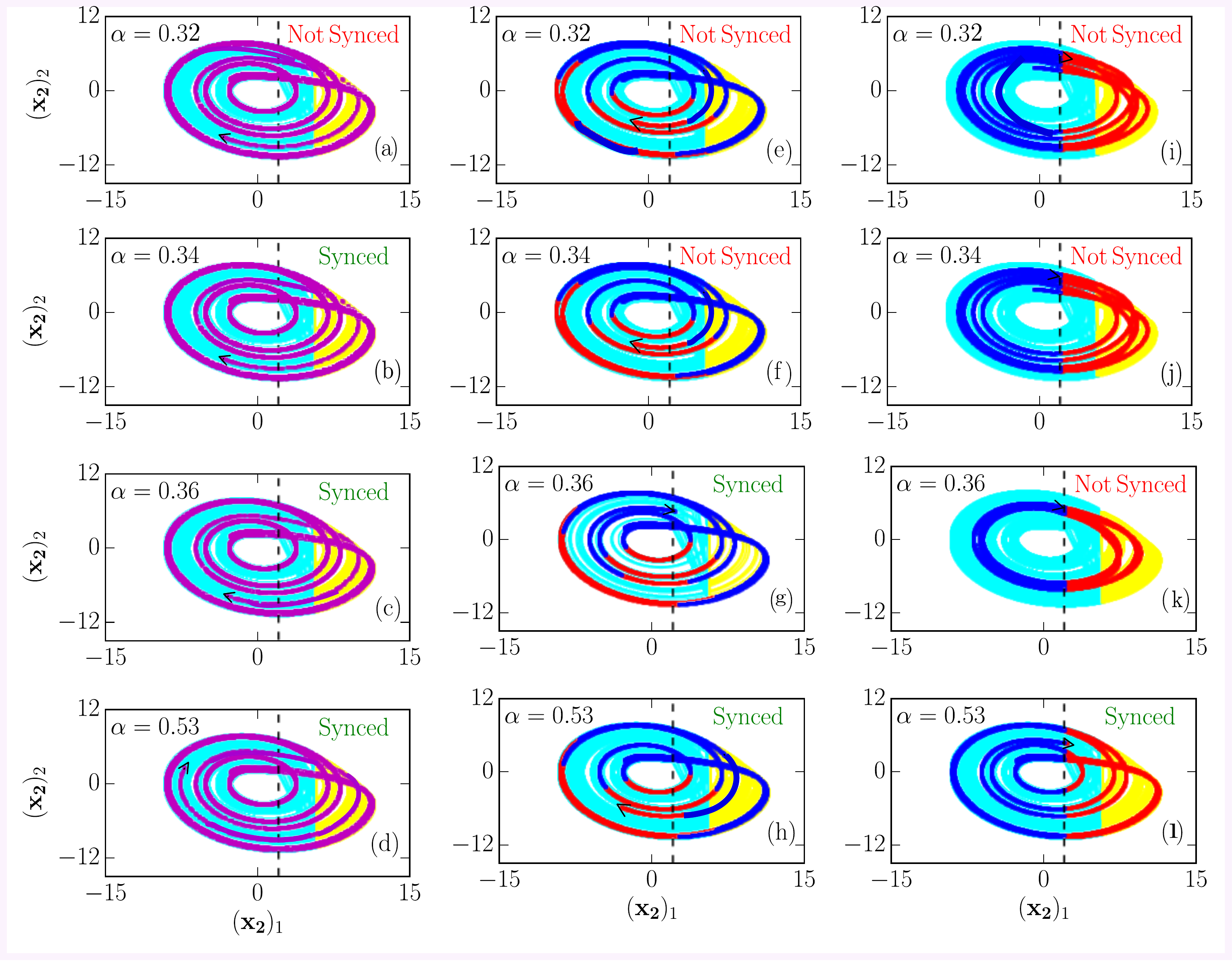}
	}%
	\caption{\textit{(Color online)} \ag{\textbf{An illustrative phase space trajectory of the driven R\"ossler oscillator under the fastest switching stochastic on-off, the transient uncoupling, and the transient stochastic uncoupling:} Synchronization starts for $x$-coupled R\"ossler oscillators using the \fss (subplots a-d), the transient stochastic uncoupling (subplots e-h), and the transient uncoupling (subplot i-l)---defined through Eq.~(\ref{eq:new_condi1}), Eq.~(\ref{eq:tus_good}), and Eq.~(\ref{eq:tus_good1}) with $x_0 = 2.1$---at $\alpha \geq 0.34, \, 0.36,$ and $0.53$ respectively (see Fig.~\ref{fig:tus}c). `Synced' and `Not Synced' respectively mark which cases are synchronized and which ones are not. The cyan and the yellow colours represent phase space regions with $\Lambda_{\rm max}<0$ and $\Lambda_{\rm max} \geq0$ respectively. The red and the blue  dots in the two dimensional projected phase spaces respectively represent the regions where coupling is on  and off. For the fastest switching stochastic on-off, since the switching of the coupling term is very fast, the blue and the red dots (that are alternately scattered all over the trajectory) are hard to depict separately without zooming in; hence, we plot the corresponding orbit in magenta.}}
	\label{fig:fig6}
\end{figure*}

Our systematic investigation with the transient stochastic uncoupling provides two significant insights about the occasional uncoupling schemes. Firstly, the conventional wisdom, that the prescription of employing a local analysis to find whether an orbit at a phase point is locally diverging away from another nearby orbit and to subsequently effect uncoupling at such points, is flawed. In fact, this flaw has been illustrated using the Lorenz and the Chen systems. \ag{This fact can be further appreciated through Fig.~\ref{fig:fig6} that showcases that the phase orbits of the driven R\"ossler oscillator in the \fss and the transient stochastic uncoupling schemes are haphazardly distributed over the locally contracting and the local non-contracting regions of the phase space but synchronization is still imparted.} Moreover, that the transient stochastic uncoupling is a successful method whenever the transient uncoupling is so, alludes to the fact that it is the autocorrelation function and not the eigenvalues of the Jacobian (or local Lyapunov exponents) that should be taken into account while trying to find the methodology of finding the optimal coupling region. After all, the autocorrelation function stays invariant when $\{s_n\}_{n=0}^{N-1}$ (used in the transient coupling)  is transformed into $\{s'_n\}_{n=0}^{N-1}$ (used in the transient stochastic uncoupling). 

Secondly, in the usual stochastic on-off uncoupling $\tau$ is kept \emph{fixed} and a \emph{constant} $p\in(0,1)$. While this definitely is quite simple prescription, we note that even the most successful version (see Table~\ref{table:2} and Fig.~\ref{fig:fss}) of the stochastic on-off uncoupling---viz., the fastest switching stochastic on-off coupling---can be bettered once these conditions are relaxed (see Fig.~\ref{fig:sub}b, e, and h). In the method of transient stochastic uncoupling, $\tau$ is no longer fixed but depends on time stochastically. Although in the case of the transient stochastic uncoupling, $p$ is not calculated explicitly, we realize that it is either $0$ or $1$ and hence is not constant. We may thus conclude that we have used a deterministic method to tweak the common stochastic method so that the resulting method of occasional uncoupling is stochastic and may supersede the fastest switching stochastic on-off coupling; all that is needed is that the method should have a favourable corresponding autocorrelation function associated with it.

\section{Discussion}
\label{sec:discussion}
In this paper, we have discussed both the deterministic and stochastic occasional uncoupling schemes that lead to the chaotic synchronization. We have seen that the \fss is superior to many of the occasional uncoupling schemes because it is applicable for many chaotic systems and not just the {coupled} R\"ossler systems. While the success of the \fss in the case of two coupled chaotic oscillators may be attributed to the effective averaged coupled dynamics characterized by an effective `renormalized' coupling parameter, the case of the transient uncoupling is far more convoluted. In fact we establish that one most probably needs to know the global information (viz., autocorrelation function of the corresponding chaotic time series) in order to find the optimal coupling region. 

In the course of our investigation, we have introduced the transient stochastic uncoupling. We must emphasize that the uncoupling method may not be a practical occasional uncoupling that could readily be employed on diffusively coupled chaotic systems. \emph{For the purpose of this paper, it is mostly an intermediate technical tool that helps us to comprehend the deterministic and the stochastic occasional uncoupling induced synchronizations through each other.} Specifically, the transient stochastic uncoupling has been invented to show how to possibly surpass the success of the \fss and to understand the importance of the autocorrelation function in the occasional uncoupling schemes.

Intriguingly, the transient stochastic uncoupling brings us back to another unsolved question mentioned at the beginning of this paper: why slow stochastic on-off works when the fast one fails? To understand how, consider Table~\ref{table:1} where we reconsider the systems studied in Section~\ref{sec:result}. The table depicts the values of $T_s$, and the values of $\alpha$ and $x_0$ chosen to employ the transient stochastic uncoupling method. The last column represents the value of $T_{\rm av}$, \emph{average switching on-off period} defined and described below.

Consider any occasional uncoupling scheme that can be described completely (see, e.g., Fig.~\ref{fig:sub}) by specifying the values of $T_{\rm sw}$ (which may be not be constant over time) and corresponding $\{S_n\}_{n\in {\mathbb N}}$. Choose a subsequence $\{\sigma_m\}_{m=1}^M$ ($M\le N$) of  $\{S_n\}_{n\in {\mathbb N}}$ such that $\{\sigma_m\}_{m=1}^M=\{1_{m'},0_{M-m'-1},1\}$ with $m'>0,\,M\ge3$; here, $1_{m'}$ means $1$ repeated $m'$ times and so on. The time elapsed in this subsequence is $\sum_{k=1}^{M-1}T_{\rm sw}(t_k)$, where $T_{\rm sw}(t_k):=t_{k+1}-t_k$ is the time elapsed between the events $S_k=0\,{\rm or }\,1$ and $S_{k+1}=0\,{\rm or }\,1$. We take the exhaustive set of all those possible subsequences, $\{\sigma_m\}_{m=1}^M$, such that given any two subsequences in the set, either the first element of one coincides with the last element of the other or there is a third subsequence in the set such that its first and last elements coincide respectively with the last element of one of the given two subsequences and the first element of the other. We calculate the times elapsed in all such subsequences and calculate their average which is $T_{\rm av}$ by definition. 

The physical meaning of $T_{\rm av}$ gets clarified on noting that $T_{\rm av}$ is analogous to $T$ in the deterministic on-off coupling method of Eq.~\ref{eq:dof} in the sense that $T_{\rm av}=T$ in the method. In passing, we also remark that $T_{\rm av}$ can easily be shown to be $\tau\sum_{k=1}^\infty k(k+1)2^{-k-1}=4\tau$ for the stochastic on-off coupling as defined in Eq.~\ref{eq:sof}, on assuming that $p$ is $0.5$ and that the coupling is turned on at time $t=0$. Thus, it is reasonable to call the stochastic on-off scheme slow or fast depending on whether $\tau$, and hence $T_{\rm av}$, is large ($\gtrsim T_s$) or small ($\ll T_s$).

\begin{table}[htpb!]
	\caption{\label{table:1} \textbf{$T_{\rm av}$~versus~$T_s$:} Comparison of the average switching on-off period ($T_{\rm av}$) with the system time scale ($T_s$) for the transient stochastic uncoupling as employed in Fig.~\ref{fig:tus}. In contrast, the numerically calculated $T_{\rm av}$ for the \fss in all the cases is equal to the analytical result, $4\tau=4h=0.04$.}
	\begin{center}
		\scalebox{1.10}{
			\begin{tabular}{lcclccrcc|} 
				\hline
				\hline
				System& & $T_s$&  $\alpha$&$x_0$ &  $T_{\rm av}$ \\
				\hline
				\hline
				\multirow{2}{3em}{R\"ossler}&& \multirow{2}{2em}{5.86}&\multirow{2}{2em}{5}& $-5.2$  &  $5.8$\\ 
				& & & &$2.1$  & $1.42$ \\ 
				\hline
				\multirow{2}{3em}{Lorenz}&& \multirow{2}{2em}{0.70}&\multirow{2}{2em}{12} & $20$  &  $0.34$ \\ 
				& & & & $30$  &  $0.30$ \\ 
				\hline
				\multirow{2}{3em}{Duffing} && \multirow{2}{2em}{2.50} &\multirow{2}{2em}{1.6} & $-1$  & $2.54$ \\ 
				& & & &$1$ &   $1.05$ \\ 
				\hline
				\hline
			\end{tabular}}
		\end{center}
	\end{table}
	Since, as seen in Table~\ref{table:1}, $T_{\rm av}$ is of the same order as the corresponding values of $T_s$, we may infer that the transient stochastic uncoupling method is akin to the slow stochastic on-off coupling in an average sense. Thus, one can not claim that the method works because the averaged dynamics, as is the case for the fast switching stochastic on-off, leads to an effective coupling parameter that results in synchronization during the continuous coupling. While we are unable to provide answer to this particular issue, which anyway is outside the scope of our present study, we think this reiterates the importance of understanding why slow stochastic on-off works. Hopefully, one will be able to answer this question in near future.
	\section*{Acknowledgements}
	The authors thank Anando G. Chatterjee, Dibakar Ghosh, Manu Mannattil, Shubhadeep Sadhukhan, Manohar K. Sharma, Sudeshna Sinha, and Saikat Sur  for fruitful discussions. S.C. gratefully acknowledges financial support from the INSPIRE faculty fellowship (DST/INSPIRE/04 /2013/000365) awarded by the INSA, India and DST, India.
	\section*{Authors contributions}
	S.C. conceptualized the study, did some related analytical calculations, and wrote the paper. A.G. did all the detailed analysis and numerical simulations as reported in the paper, made all the figures, and helped in writing the paper. 

	\appendix
	\section{Effectiveness of the \fss}
	\label{sec:appendix}
	Since almost all of the diffusive occasional uncoupling schemes are tested primarily with the coupled R\"ossler oscillators~\cite{roessler76}, we use the same system to see the effectiveness of the fastest switching stochastic on-off coupling to begin with. The corresponding explicit form of Eq.~\ref{eq:1}b for the driven R\"ossler oscillator, thus, is:
	\begin{subequations}
		\begin{eqnarray}
			\dot{x}_2 &=&  -y_2-z_2 + \alpha \chi_{(\xi, q)}(t) \cdot (x_1- x_2), \\
			\dot{y}_2 &=&x_2+ay_2,\\
			\dot{z}_2 &=&b + z_2(x_2-c).
		\end{eqnarray}\label{eq:ross}
	\end{subequations}
	Here, $a = b = 0.2$ and $c = 5.7$. We note that the subsystems are $x$-coupled, i.e., $\textsf{C}_{11}=1$ is the only non-vanishing element of the coupling matrix $\textsf{C}$. {It may be mentioned that the average inter-peak length of the $x$-time-series, which can be taken as the system's innate timescale, is $T_s \approx 5.86$ for R\"ossler oscillator.} For the fastest switching stochastic on-off coupling, we work with $\tau = 0.01\ll T_s$.
	
\begin{figure}[h]
		\includegraphics[width=30 cm,height=6.8cm, keepaspectratio]{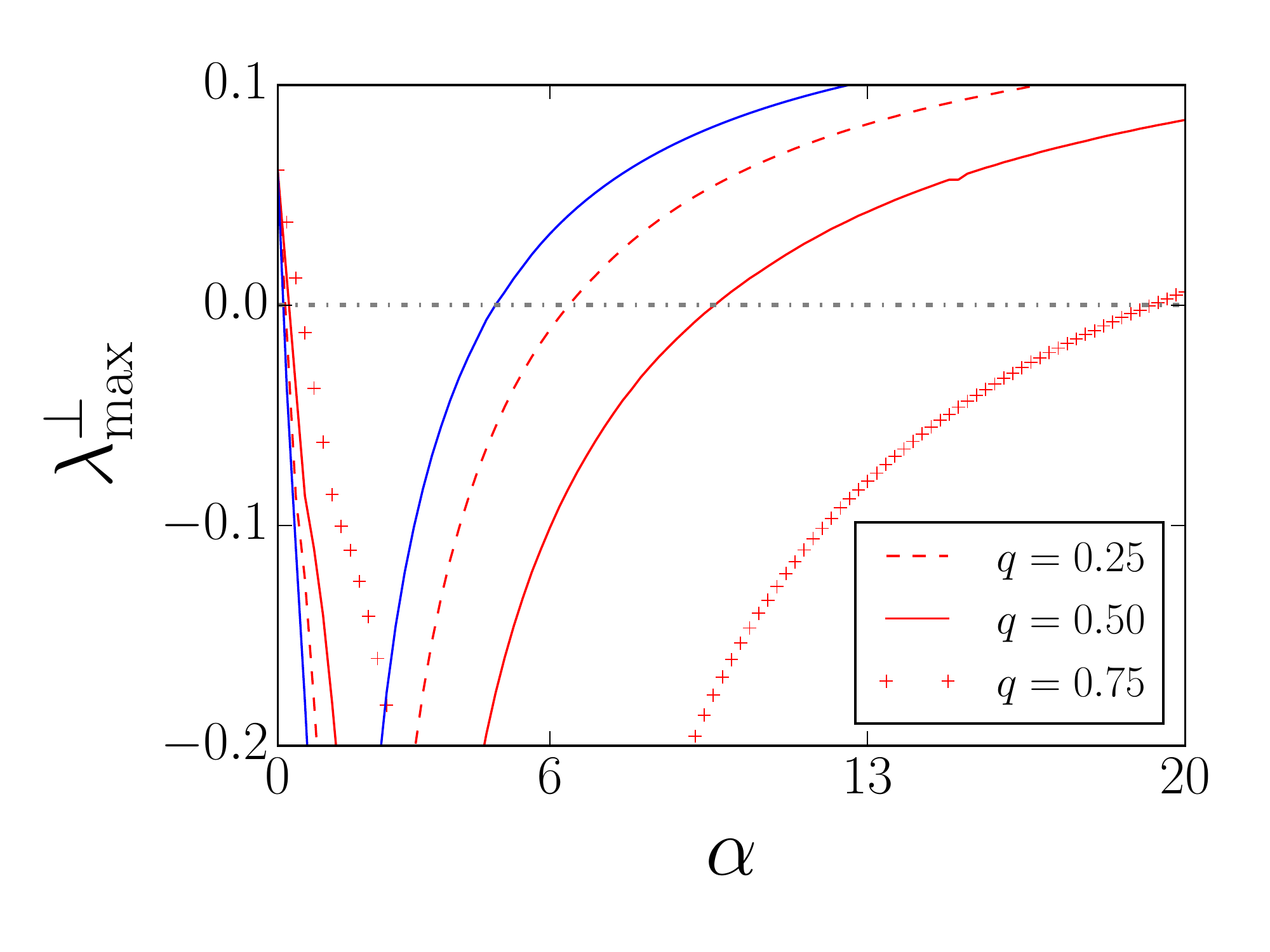}
		\caption{\textit{(Color online)} \textbf{The \fss is more effective than the continuous coupling:} The maximum conditional Lyapunov exponent ($\lambda^{\perp}_{\rm max}$) is plotted with the coupling strength ($\alpha$) for the $x$-coupled R\"ossler oscillators using the continuous coupling (blue curve) and the \fss {(red dashed curve and solid curve are respectively for $q = 0.25$ and $q = 0.5$, and the red curve with $`+'$ is for $q =0.75$)}. Since $\lambda^{\perp}_{\rm max}<0$ for comparatively higher values of $\alpha$, the \fss appreciably extends the range of $\alpha$ where synchronized state are effected.}
		\label{fig:cle_ccs_rcs}
	\end{figure}

	The negativity of the maximum conditional Lyapunov exponent ($\lambda^{\perp}_{\rm max}$)~\cite{pea1997} is a signature of a synchronized state when two subsystems are coupled diffusively. In Fig.~\ref{fig:cle_ccs_rcs}, we see that while continuously coupled R\"ossler oscillators fails to lead to the robust synchronized state for values of $\alpha \gtrsim 4.4$, on implementing the \fss, the oscillators can evolve synchronously up to $\alpha\approx 5.86,\, 8.8,$ and $17.6$ for $q = 0.25, \,0.5,$ and $0.75$ respectively. 
	In Table~\ref{table:2}, we tabulate similar advantage of the \fss in effecting synchronization in some other well-known chaotic systems---the simplest cubic chaotic flow~\cite{malasoma00}, the Halvorsen's cyclically symmetric attractor~\cite{sprott03}, the Lorenz oscillator~\cite{lor1963}, and the Chen oscillator~\cite{chen1999}: We note that the \fss is capable of inducing synchrony in these systems even when they do not synchronize for the entire corresponding ranges of parameter values and different kinds of coupling schemes. Evidently, the success of the \fss is quite impressive.
	\begin{sidewaystable*}[!htbp]
		\begin{center}
			\vspace*{18 cm}
			\caption{\textbf{The \fss is the most effective occasional uncoupling method:} The table below compares the effectiveness of different occasional uncoupling schemes of various oscillators that are chaotic to varying degree---the maximum Lyapunov exponents of the R\"ossler oscillator, the simple cubic chaotic flow, the Halvorsen's cyclically symmetric attractor, the Lorenz system, and the Chen system are $0.07$, $0.08$, $0.789$, $0.90$, and $2.02$ respectively~\cite{sprott03}. Also, refer to Fig.~\ref{fig:fss}.}
			\scalebox{0.9}{
				\begin{tabular}{ccccccccc}
					\hline
					\hline                       
					& Name of the & Range of $\alpha$ & Range of $\alpha$ & Sporadic & On-off & Stochastic & Transient & Fastest
					\\
					Serial No. & oscillators   & over which & investigated & on-off~\cite{stojanovski96} &  coupling\footnote{In the case of small $T\,(T\ll T_s)$, the on-off coupling scheme can be replaced by the continuous coupling with an effective coupling strength $\alpha_{\rm eff} = \alpha \theta$ and hence can be as successful as the \fss.}~\cite{cqh09}& coupling~\cite{jeter15} & uncoupling~\cite{sch15} & \quad switching
					\\
					&  &the continuous coupling &  &$(\Delta t \in [0.02, 1.0])$ & & & (following Eq.~\ref{eq:tus_condi}) & \quad stochastic
					\\
					& &synchronizes&  &  & & & & on-off \\
					& & & & & & & &  $(\tau = 0.01, q = 0.5)$
					\\
					[0.1ex]
					\hline
					\hline
					& R\"ossler~\cite{roessler76} & \multirow{3}{*}{$[0.14, 4.4]$}& \multirow{3}{*}{$[4.4,15]$} &  & \textbf{Synced} for &\textbf{Synced} for & \textbf{Synced} for  & \textbf{Synced} for 
					
					\\
					& $\dot{x}_1 = -y_1-z_1,$  &  & &Not synced &$\alpha \in [4.4, 15]$ &$\alpha \in [4.4, 15]$ &$\alpha \in [4.4, 15]$ & $\alpha \in [4.4, 9.45]$
					\\
					$1$ & $\dot{y}_1 =x_1+ 0.2y_1,$   & & &  &($T = 3$, & $(\tau = 2$, &$(x_{10} = 1.20$, &
					\\
					& $\dot{z}_1 = 0.2 + z_1(x_1-5.7).$   & & & &$\theta = 0.5$) &$p = 0.5$) &$\Delta = 4.16)$ & 
					\\
					& (Using $x$-coupling)&  & & & & &  &
					\\
					\hline
					& Simplest cubic  &  \multirow{3}{*}{$[0.16, 1.02]$}&\multirow{3}{*}{$[1,10]$} &  & Not synced & Not synced & Not synced &\textbf{Synced} for 
					\\
					& chaotic flow~\cite{malasoma00}   & & & &$(T \in [1, 4]$,  &$(\tau = [1, 4]$, & $(x_{10} = 0,$ & $\alpha \in [1, 2.35]$
					\\
					$2$& $\dot{x}_1 = y_1,$ &   & &Not synced &$\theta \in [0.05, 0.95])$ &$p \in [0.05, 0.95])$ &$\Delta = 1)$ &
					\\
					& $\dot{y}_1 =z_1,$ &  & & & &  & &
					\\
					& $\dot{z}_1 = -2.028 z_1 + x_1 y_1^2 -x_1.$  & & & & & & &
					\\
					& (Using $x$-coupling)&  & & & & & &
					\\
					\hline
					& Halvorsen's cyclically  & \multirow{3}{*}{$[7.52, 47.00]$}& \multirow{3}{*}{$[47, 70]$} &  & Not synced & Not synced & Not synced & \textbf{Synced} for 
					\\
					& symmetric attractor~\cite{sprott03}   & & & &$(T \in [1, 4],$ &$( \tau \in [1,4],$ & $(x_{10} = 0,$&$ \alpha \in [47, 70]$
					
					\\
					$3$ & $\dot{x}_1 = -1.27x_1 -4y_1-4z_1-y_1^2,$ & & &Not synced &$\theta \in [0.05, 0.95])$ &$p \in [0.05, 0.95])$ & $\Delta  =2)$ &
					\\
					& $\dot{y}_1 =-1.27y_1 -4z_1-4x_1-z_1^2,$  & & & & & &  &
					\\
					& $\dot{z}_1 = -1.27z_1 -4x_1-4y_1-x_1^2.$ & & & & &  & &
					\\
					
					& (Using $x$-coupling) &  & & & & & &
					\\
					\hline
					& Lorenz~\cite{lor1963}  & \multirow{3}{*}{$[1.50, 6.50]$} & \multirow{3}{*}{$[6.5,100]$}&  & Not synced & Not synced& \textbf{Synced} for & \textbf{Synced} for 
					\\
					& $\dot{x}_1 =  10 (y_1 - x_1),$   & & &Not synced & $(T \in [1, 4],$&$( \tau \in [1, 4],$ &$\alpha \in [30.5, 100]$ &$\alpha \in [6.5, 13]$
					\\
					$4$ & $\dot{y}_1 =-x_1 z_1 + 28 x_1-y_1,$ & & & & $\theta \in [0.05, 0.95])$& $p \in [0.05, 0.95])$ & $(x_{30} = 25$, &
					\\
					& $\dot{z}_1 = x_1 y_1 -(8/3) z_1.$ & & & & & & $\Delta = 1$)&
					\\
					& (Using $z$-coupling)&   & & & & & & 
					\\
					\hline
					& Chen~\cite{chen1999} & \multirow{3}{*}{$[4.54, 21]$} & \multirow{3}{*}{$[21,100]$}& & Not synced  & Not synced & \textbf{Synced} for  & \textbf{Synced} for 
					\\
					
					& $\dot{x}_1 = 35(-x_1  + y_1),$  & & & Not synced& $(T \in [1, 4]$,& $(\tau \in [1,4],$&$\alpha \in [40, 100]$ &$ \alpha \in [21, 46.1]$ 
					\\
					$5$ & $\dot{y}_1 = -7x_1 -x_1 z_1+ 28 y_1,$  & & &  & $\theta \in [0.05, 0.95])$& $p \in [0.05, 0.95])$& $(x_{30} = 26.5$, &
					\\
					& $\dot{z}_1 = x_1 y_1 -3 z_1.$ & & & & &  & $\Delta = 5$)&
					\\
					
					& (Using $z$-coupling) & & & & & & &
					\\
					[1ex]
					\hline
					\hline
				\end{tabular}}
				\label{table:2}
			\end{center}
		\end{sidewaystable*}

		\begin{figure*}[htbp!]
			{
				\hspace*{0.3 cm}
				\includegraphics[width=74cm,height=20cm, keepaspectratio]{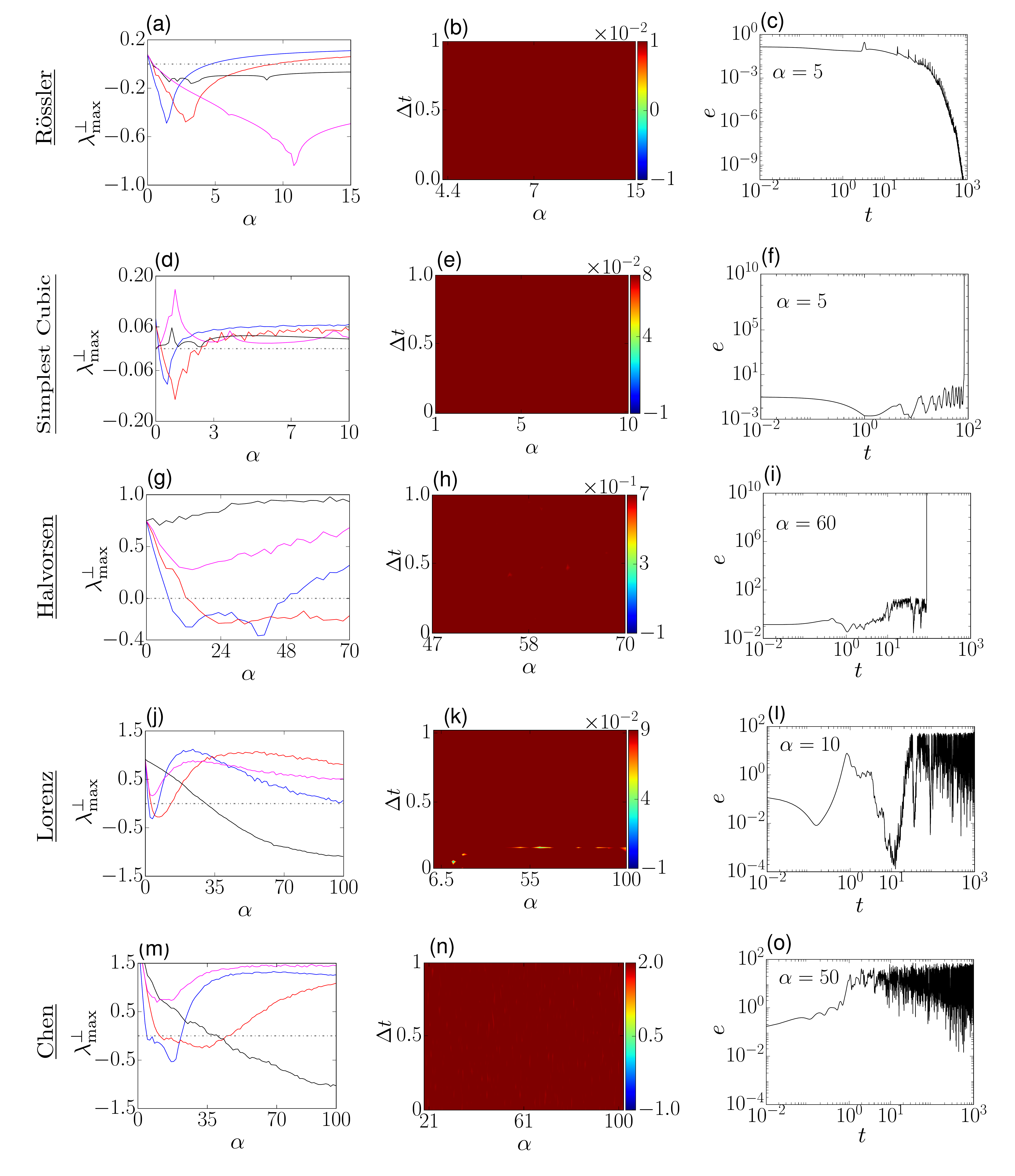}
			}%
			\caption{\textit{(Color online)} \textbf{The \fss is the most effective occasional uncoupling method:}
				The five rows from the top to the bottom correspond to different coupled oscillators, viz., the R\"ossler system, the simplest cubic chaotic flow, the Halvorsen's cyclically symmetric attractor, the Lorenz oscillator, and the Chen system respectively. The \fss---along with four different occasional coupling schemes---the sporadic coupling, on-off coupling, the transient uncoupling, and the stochastic on-off coupling---is employed on these diffusively coupled chaotic oscillators. In the leftmost subplot of each row (i.e., subplots a, d, g, j, and m), the blue, the black, the magenta, and the red curves respectively correspond to the continuous coupling, the transient uncoupling (with coupling regions as tabulated in Table~\ref{table:2}), the on-off coupling (with $T=3$ and $\theta=0.5$), and the \fss (with $\tau=h=0.01$ and $q=0.5$). The mesh-grid plots in the middle column (i.e., subplots b, e, h, k, and n) are for the sporadic coupling applied on the respective system with $\Delta t$ varying from $0$ to $1$ in steps of $0.02$. The color bars in the mesh-grid plots quantify the values of $\lambda_{\rm max}^\perp$ whose positive values imply desynchronization in all the five corresponding subplots. The rightmost plot in each row (i.e, subplots c, f, i, l, and o) correspond to the (slow) stochastic on-off coupling with $(\tau, p) = (2, 0.5)$ and is plotted for a single value of $\alpha$ chosen within the range given in Table~\ref{table:2} for illustrating the conclusions of  table.}
			\label{fig:fss}
		\end{figure*}
		%
		
		%
		%
		The success of the fast switching stochastic on-off coupling in the case of \emph{two} diffusively coupled oscillators is easy to understand. To this end, we write the general form of equation of motion of the driven oscillator using the mentioned scheme in the following form:
		\begin{equation}
			\frac{d\mathbf{x}_2}{dt} = \mathbf{F(x}_2) + \alpha  \chi_{(\xi, q)} (t)
			\sf{C}\cdot(\mathbf{x}_1 - \mathbf{x}_2).
		\end{equation}
		After each time-step $h$, we can formally write the solution of the above equation as
		\begin{eqnarray}
			\label{eq:averaging_eqn}
			\mathbf{x}_2(t+h) = &&\mathbf{x}_2(t) + \int_{t}^{t+h}dt\,
			\mathbf{F(x}_2)\nonumber\\  \phantom{\mathbf{x}_2(t+h) = }&&+ \alpha \int_{t}^{t+h} dt\, \chi_{(\xi, q)}
			(t)  \sf{C}\cdot(\mathbf{x}_1 - \mathbf{x}_2).
		\end{eqnarray}
		$h$ is chosen in such a manner that $h\ll T_s$, the corresponding system's time scale, so that the function $\mathbf{F}(\mathbf{x}_2)$ and
		$ \sf{C}\cdot(\mathbf{x}_1 - \mathbf{x}_2)$, remain practically constant over the time interval $h$. Consequently, it follow from Eq.~\ref{eq:new_condi} and Eq.~\ref{eq:averaging_eqn} that
		\begin{equation}
			\mathbf{x}_2(t+h) \approx \mathbf{x}_2(t) + h \mathbf{F(x}_2) +
			h{\alpha}_{\rm eff}  \sf{C}\cdot(\mathbf{x}_1 - \mathbf{x}_2),
		\end{equation}
		where ${\alpha}_{\rm eff} := \alpha (1-q) < \alpha$, since $q \in (0,1)$. It means that whatever is the state of synchrony of the occasionally uncoupled system is at a given value of $\alpha$, it should be effectively in the same state as that of the continuously coupled system for $\alpha_{\rm eff}$.

		For R\"ossler oscillator, we choose $h\ll T_s \approx 5.86$ and employ the \fss with $(\tau,q )= (0.01, 0.25), (0.01, 0.5),$ and $ (0.01, 0.75)$. It is clear from Fig.~\ref{fig:cle_ccs_rcs} that the upper threshold of synchronization, i.e., the coupling strength at which $\lambda_{\max}^\bot=0$, is pushed from $\alpha \approx 4.4$ (continuously coupled case) to $\alpha \rightarrow4.4/(1-q) \approx 5.86, \, 8.8,$ and $17.6$ for $q = 0.25, \, 0.5,$ and $0.75$ respectively. In other words, $\alpha\approx 5.86, \, 8.8,$ and $17.6$ in the case of the fastest switching stochastic on-off uncoupling for $q = 0.25, \, 0.5,$ and $0.75$ respectively are equivalent to the continuously coupled system at $\alpha=\alpha_{\rm eff}=5.86(1-0.25)=8.8(1-0.5)=17.6(1-0.75)=4.4$.
		
		Needless to say, owing to the general mechanism detailed above, the \fss is widely applicable in yielding a synchronized state for two coupled oscillators unlike many of the occasional uncoupling schemes. {{However, this uncoupling scheme may fail to give rise to synchronized states when one goes beyond a system of two coupled oscillators, e.g., a network of oscillators}~\cite{jeter14}}. Intriguingly enough, the stochastic on-off uncoupling with slow switching (in which $\tau$ and $T_s$ have same order of magnitudes) leads to synchronized states~\cite{jeter15} in such networks.
\bibliographystyle{unsrt}
\bibliography{Ghosh_Chakraborty_manuscript.bib}
\end{document}